# Synthesizing cross-design evidence and cross-format data using network meta-regression

Tasnim Hamza[1,2], Konstantina Chalkou[1,2], Fabio Pellegrini[3], Jens Kuhle[4], Pascal Benkert[5], Johannes Lorscheider[4], Chiara Zecca[6,7], Cynthia P Iglesias-Urrutia [8], Andrea Manca[9], Toshi A. Furukawa[10], Andrea Cipriani[11], Georgia Salanti[1].

[1]Institute of Social and Preventive Medicine, University of Bern, Bern, Switzerland.
[2]Graduate School for Health Sciences, University of Bern, Switzerland.
[3]Biogen Digital Health, Biogen Spain, Madrid, Spain.
[4]Department of Neurology, University Hospital Basel and Departments of Biomedicine and Clinical Research, University Hospital Basel, University of Basel, Basel, Switzerland.
[5]Department of Clinical Research, University of Basel, Basel, Switzerland.
[6]Multiple Sclerosis Center, Neurocenter of Southern Switzerland, EOC, Lugano, Switzerland
[7]Faculty of biomedical Sciences, Università della Svizzera Italiana, Lugano, Switzerland.
[8]Department of Health Sciences, University of York, York, United Kingdom.
[9]Centre for Health Economics, University of York, York, United Kingdom.
[10]Department of Health Promotion and Human Behavior, and Department of Clinical Epidemiology, Graduate School of Medicine/School of Public Health, Kyoto University, Kyoto, Japan.
[11]Department of Psychiatry, University of Oxford, Oxford, United Kingdom.
[12]Oxford Health NHS Foundation Trust, Warneford Hospital, Oxford, United Kingdom.

**Funding**

TH, KH, CI, AM and GS are funded by the European Union's Horizon 2020 research and innovation programme under grant agreement No 825162. AC is supported by the National Institute for Health Research (NIHR) Oxford Cognitive Health Clinical Research Facility, by an NIHR Research Professorship (grant RP-2017-08-ST2-006), by the NIHR Oxford and Thames Valley Applied Research Collaboration, and by the NIHR Oxford Health Biomedical Research Centre (grant BRC-1215-20005). The views expressed are those of the authors and not necessarily those of the UK National Health Service, the NIHR, or the UK Department of Health. TAF reports grants and personal fees from Mitsubishi-Tanabe, personal fees from MSD, personal fees from SONY, grants and personal fees from Shionogi outside the submitted work. In addition, TAF has a patent 2020-548587concerning smartphone CBT apps pending, and intellectual property for Kokoro-app licensed to Mitsubishi-Tanabe. CZ holds a grant from Ente Ospedaliero Cantonale for senior researchers.

## Conflict of interest

FP is an employee of and holds stocks/stock options in Biogen. JL has received research support from Innosuisse – a Swiss innovation agency, Biogen. and Novartis and received speaker honoraria and/or compensation for serving on advisory boards from Roche, Teva, and Novartis. Ente Ospedaliero Cantonale (employer) received compensation for Chiara Zecca's speaking activities, consulting fees, or research grants from Abbvie, Almirall, Biogen Idec, Bristol Meyer Squibb, Genzyme, Lundbeck, Merck, Novartis, Teva Pharma, and Roche. Other authors have no conflicts of interest to declare relevant to the content of this article.

## Acknowledgment

We are very grateful to Suvitha Subramaniam for her great help extracting data from the Swiss Multiple Sclerosis Cohort registry for this project.

## Abstract (250 words)

In network meta-analysis (NMA), we synthesize all relevant evidence about health outcomes with competing treatments. The evidence may come from randomized controlled trials (RCT) or non-randomized studies (NRS) as individual participant data (IPD) or as aggregate data (AD). We present a suite of Bayesian NMA and network meta-regression (NMR) models allowing for cross-design and cross-format synthesis. The models integrate a three-level hierarchical model for synthesizing IPD and AD into four approaches. The four approaches account for differences in the design and risk of bias in the RCT and NRS evidence. These four approaches variously ignoring differences in risk of bias, using NRS to construct penalized treatment effect priors and bias-adjustment models that control the contribution of information from high risk of bias studies in two different ways. We illustrate the methods in a network of three pharmacological interventions and placebo for patients with relapsing-remitting multiple sclerosis. The estimated relative treatment effects do not change much when we accounted for differences in design and risk of bias. Conducting network meta-regression showed that intervention efficacy decreases with increasing participant age. We re-analysed a network of 431 RCT comparing 21 antidepressants, and we did not observe material changes in intervention efficacy when adjusting for studies' high risk of bias. In summary, the described suite of NMA/NMR models enables inclusion of all relevant evidence while incorporating information on the within-study bias in both observational and experimental data and enabling estimation of individualized treatment effects through the inclusion of participant characteristics.

# 1 Introduction

Network meta-analysis (NMA) is a widely used tool to synthesise the available evidence that may vary in design and format (1–3). Evidence may come either from a randomized controlled trial (RCT) or a non-randomized study (NRS); as either individual participant data (IPD) or aggregate data (AD). As heterogeneity is a common attribute of evidence synthesis, many published comparative effectiveness reviews account for covariates that modify the treatment effect in a network meta-regression (NMR) (4,5). The effect of study-level covariates can be modelled using only AD, while IPD is needed to adjust for patient-level covariates to avoid aggregation bias (6). The inclusion of these participant characteristics also enables estimating individualized treatment effects.

Matching-Adjusted Indirect Comparison (7) and simulated treatment comparison methods (8) have been used to combine evidence from IPD and AD using reweighting techniques and regression models, respectively, to adjust for effect modifiers. However, this adjustment needs to be done separately for each treatment comparison and requires IPD for at least one of each treatment comparison. It was shown that simulated treatment comparison methods perform poorly (9,10). Jansen proposed combining IPD and AD in an NMA by integrating the underlying IPD distribution of the AD studies (11). The method was applied initially to binary outcomes and extended to other data types (12). The three-level hierarchical model extends the standard NMR model combining IPD and AD by introducing a new level differentiating between the two formats (11,13–15).

While most published NMAs only synthesize RCT evidence, there is growing interest incorporating non-randomized or real-world evidence in these analyses (16,17). The inclusion of evidence from NRS has many potential advantages, such as better reflected clinical practice realities; the data in follow-up studies are collected over relatively long time periods; and finally, NRS are essential when RCTs are less feasible (eg, in rare conditions). While RCT evidence is considered to be of lower risk of bias when compared with NRS, a Cochrane review found little evidence that RCTs and NRSs provide different estimates of treatment effect (18). Also, many empirical studies have identified different types of bias possibly present in many RCTs. For example, Schulz et al. (19) found that RCTs with inadequate allocation concealment or lack of blinding tended to exaggerate the estimated treatment effect and provide biased results. Similarly, Chalmers et al. (20) showed major differences between treatment and control effects in unblinded trials, as well as trials lacking proper randomisation when compared with double-blinded studies. Wood et al. (21) found that the treatment effect estimates of subjective outcomes (outcomes are dependent on judgment from an assessor or patient-reported) were exaggerated for studies with poor allocation concealment or lack of blinding.

Several methods have been proposed for combining various designs in NMA contexts. Three methods have been presented to synthesize RCT and NRS evidence (22,23): combining studies of different design and ignoring their differences in the naïve approach; using NRS evidence to construct penalized treatment effect priors; and adding a new level to reflect the design differences in a three-level hierarchical model. The latter requires the network to include many studies on each design which might not be the case for most NMAs (23). Dias et al. (24) presented an NMA model that adjusts for the within-study risk of bias (RoB) by adding a bias indicator. The bias indicator was assigned a binary value of 0 for low RoB studies; 1 for high RoB studies; and a uniform distribution for studies with unclear RoB. Verde (25) proposed to model the unadjusted and adjusted relative treatment effect simultaneously using a bimodal normal distribution. The model was developed for pairwise meta-analysis.

We extend the two RoB adjustment methods described above by accounting for the uncertainty in each RoB judgment in Dias et al.'s model and by drawing from Verde's approach into NMA (24,25). Then we build a Bayesian cross-NMA/NMR model by integrating the approaches that combine RCT and NRS evidence into the three-level hierarchical model, which combines IPD and AD. This model enables estimating treatment effects among specific subgroups of patients through the inclusion of participant characteristics.

This work has been done within the HTx Horizon 2020 project. HTx is supported by the European Union, lasting for 5 years from January 2019. The main aim of HTx is to create a framework for the Next Generation Health Technology Assessment (HTA) to support patient-centered, societally oriented, real-time decision-making on access to and reimbursement for health technologies throughout Europe.

## 2 Examples

We analysed two networks of interventions: one of pharmacological agents in relapsing-remitting multiple sclerosis (RRMS) and another of antidepressant treatments (Figure 1). In both examples, RoB judgements were formulated using the Cochrane RoB tool 1 (26).

**Relapsing-Remitting Multiple Sclerosis (RRMS) drugs network:** The agents to manage RRMS were compared in systematic reviews of RCTs and NMAs (27,28). We contribute to the methodological literature by analysing the IPD and AD from five RCTs (29–33) and the Swiss Multiple Sclerosis Cohort (SMSC) (34).

We defined the inclusion criteria for patients from the SMSC to be consistent with the RCTs' criteria. We only included people from the SMSC with RRMS treated with any of the three active agents shown in Table 3. Compared with available RCTs, individuals in the SMSC are followed for longer. To avoid immortal time bias, we specified the length and the start of follow-up for each individual (35,36). Since two years was the typical duration of the RCTs we included, we defined cycles of length of two years from when a patient initiated a treatment in SMSC; we recorded their outcome during these two years of follow-up.

To investigate the effectiveness of the treatments in subgroups of people, we explored whether age at the time of treatment initiation modifies the treatment effect. Individuals with RRMS have flare-ups of relapses or symptoms; between these flare-ups, they are free of symptoms (37). Our outcome of interest is relapse at two years of follow-up. We use the odds ratio (OR) to compare treatments. When OR of treatment A vs B ($OR_{AB} = odds_A/odds_B$) is less than 1, treatment A is more effective than treatment B.

Figure 1a, Table 3 and Appendix Table 1-3 summarize the data available, their format and the RoB in each study.

**Antidepressants:** Our dataset includes AD from 431 RCTs (263 at moderate RoB and 168 at low RoB) comparing 21 antidepressants and placebo (38). The outcome of interest is response to treatment defined as 50% reported reduction in depression symptoms. In the original article, the authors performed a sensitivity analysis by including only low RoB studies in their analysis (38). We re-analysed their dataset by controlling the impact of information from studies at different levels of RoB. (The dataset is available at https://data.mendeley.com/datasets/83rthbp8ys/2.)

# 3  Methods

We review existing NMR models to combine different data formats—IPD or AD—in this section. We then extend these models by combining the evidence from RCT and NRS in four different ways. The models we later introduce are implemented in a new R package called *crossnma* available on GitHub (https://github.com/htx-r/crossnma), as is the R code for the analysis of both examples and the antidepressant dataset (https://github.com/htx-r/crossnma-theoretical-paper-analysis).

## 3.1 Synthesizing cross-format data: individual participant data (IPD) and aggregate data (AD)

To combine IPD and AD data into the three-level hierarchical model of network meta-regression, we divided the model into three parts; in the first two parts, the model is set for IPD and AD separately. Next, we present how we combined the evidence from both parts. We describe all models assuming binary outcomes; however, they can be adapted easily to other outcome types, such as time-to-event data, as described by Saramago et al. (39). The NMA models are simply the NMR models without covariate terms.

**Part I: Network meta-regression (NMR) model for IPD studies**

Assuming $y_{ijk}$ is a binary outcome of participant $i$ in study $j$ under treatment arm $k$, we place a Bernoulli distribution for $y_{ijk}$ with a probability of an event to occur $p_{ijk}$. This probability is then linked to the control/treatment effect via a logistic transformation. The study-specific baseline effect $u_{jb}$ is the log-odds in the reference treatment $b$ in that study. The treatment effect $\delta_{jbk}$ represents the log odds ratio of treatment $k$ relative to the reference treatment $b$. Both effects $\delta_{jbk}$ and $u_{jb}$ are defined when the participant and mean covariates equal zero.

To estimate subgroup-specific treatment effects, we consider the covariate effect by adding the following three parameters (i) a regression coefficient, $\beta_{0j}$, which captures the prognostic effect of the covariate in study $j$; (ii) a between-study regression coefficient, $\beta^B_{1,jbk}$, which quantifies the interaction between the relative treatment effect and the mean covariate value across studies; and (iii) a within-study regression coefficient, $\beta^W_{1,jbk}$, which models the treatment-covariate interaction effect at the individual level. The two coefficients $\beta_{0j}$ and $\beta^W_{1,jbk}$ are estimated using the participant-level covariate $x_{ijk}$, while $\beta^B_{1,jbk}$ requires only the study mean covariate ($\bar{x}_j$) that is often reported in the publication. The term $\beta^B_{1,jbk} - \beta^W_{1,jbk}$ quantifies the discrepancy among the between- and the within-covariate estimates or the aggregation bias. In the following, we summarize the likelihood and the parametrization of the model in IPD studies:

$$y_{ijk} \sim Bernoulli(p_{ijk})$$

$$\text{Logit}(p_{ijk}) = \begin{cases} u_{jb} + \beta_{0j} x_{ijk} & \text{if } k = b \\ u_{jb} + \delta_{jbk} + \beta_{0j} x_{ijk} + \beta^W_{1,jbk} x_{ijk} + (\beta^B_{1,jbk} - \beta^W_{1,jbk}) \bar{x}_{j.} & \text{if } k \neq b \end{cases}$$

Where $j = 1, \ldots, ns_{IPD}$, and $ns_{IPD}$ is the total number of IPD studies.

**Part II: NMR model for AD studies**

We model the published information from each AD study next. For each treatment $k$ in study $j$, we place a binomial distribution for the corresponding number of events $r_{jk}$ with sample size $n_{jk}$ and probabilities of the event to occur $p_{.jk}$.

$$r_{jk} \sim Bin(p_{.jk}, n_{jk})$$

$$Logit(p_{.jk}) = \begin{cases} u_{jb} & \text{if } k = b \\ u_{jb} + \delta_{jbk} + \beta^B_{1,jbk} \bar{x}_j & \text{if } k \neq b. \end{cases}$$

We incorporate the study-level covariate effect by adding only $\beta^B_{1,jbk} \bar{x}_j$. Here, $j = 1 + ns_{IPD}, \ldots, ns_{IPD} + ns_{AD}$ where $ns_{AD}$ is the total number of AD studies.

**Part III: Combine the evidence from IPD and AD**

We combine the relative treatment effects and the between-study regression coefficients from IPD and AD parts via an exchangeable model

$$\delta_{jbk} \sim N(d_{Ak} - d_{Ab}, \tau^2), \qquad \beta^B_{1,jbk} \sim N(B^B_{1,Ak} - B^B_{1,Ab}, \tau^2_B),$$

where $j = 1, \ldots, ns_{IPD} + ns_{AD}$.

The within-study regression estimates from only IPD studies ($j = 1, \ldots, ns_{IPD}$) are synthesized as

$$\beta^W_{1,jbk} \sim N(B^W_{1,Ak} - B^W_{1,Ab}, \tau^2_W).$$

Here, $A$ represents the reference treatment in the whole network; therefore, $d_{AA}, B^W_{1,AA}, B^B_{1,AA} = 0$.

Alternatively, a fixed-effect model can be assumed

$$\delta_{jbk} = d_{Ak} - d_{Ab}, \qquad \beta^B_{1,jbk} = B^B_{1,Ak} - B^B_{1,Ab}, \qquad \beta^W_{1,jbk} = B^W_{1,Ak} - B^W_{2,Ab}.$$

We summarize the model assumptions in Table 2.

We assumed minimally informative priors for $u_{jb}, \beta_{0j} \sim N(0, 10^2)$ and also for the basic parameters $B^W_{1,Ak}, B^B_{1,Ak}, d_{Ak} \sim N(0, 10^2)$. For all heterogeneity parameters, we assigned a uniform distribution $\tau, \tau_B, \tau_W \sim Unif(0,2)$ which allows for difference of log-odds ratios of 2 across trials in the treatment and the covariate effect. This change is adequately large; hence, the given prior can be considered sufficiently vague.

In all models we present with random treatment effects, we accounted for correlations induced by multi-arm studies using a multivariate distribution as in the standard NMA methods (2). In Appendix 2, we describe how we accounted for multi-arm studies in bias-adjusted model 2.

## 3.2 Synthesizing cross-design data: Randomized trials and observational data

The model we described in Section 3.1 can be applied to RCT or NRS studies, separately. Next, we describe four different approaches to combine evidence from RCTs and NRSs into the model from Section 3.1.

### *3.2.1 Unadjusted network meta-regression*

Using the simplest approach, we integrate the NRS evidence into the RCT model without differentiation between the two designs. Technically, this means we only need to expand the index of study $j$ to involve both study designs. For IPD, it becomes $j = 1, \ldots, ns_{IPD,RCT} + ns_{IPD,NRS}$ and in AD part, $j = ns_{IPD,RCT} + ns_{IPD,NRS} + 1, \ldots, ns_{IPD,RCT} + ns_{IPD,NRS} + ns_{AD,RCT} + ns_{AD,NRS}$.

### *3.2.2 Using non-randomized studies (NRS) to construct priors for the treatment effects*

Using NRS evidence to construct priors for the treatment effects in the RCT model is a two-step approach. In the first step, the (network) meta-regression—with only NRS data—estimates the relative treatment effects with posterior distribution of mean $\tilde{d}_{Ak}^{NRS}$ and variance $V_{Ak}^{NRS}$. In the second step, the posteriors of NRS results—accounting for possible confounders—are then used as priors for the corresponding basic parameters in the RCT model; $d_{Ak} \sim N(\tilde{d}_{Ak}^{NRS}, V_{Ak}^{NRS})$. Treatment effects not observed in NRS are given vague priors (see part III of Section 3.1).

We can control the potential dominance of NRS evidence (eg, because of the large sample size) on the RCT model by either shifting the NRS means with a bias term $\varsigma$ or by dividing the variance in the prior distribution with a common inflation factor $w$, $0 < w < 1$; $d_{Ak} \sim N(\tilde{d}_{Ak}^{NRS} + \varsigma, V_{Ak}^{NRS}/w)$. When $w = 1$, NRS evidence is used at face value and when $w \approx 0$, NRS evidence is ignored.

### *3.2.3 Bias-adjusted model 1*

We incorporate judgments about study risk of bias in bias-adjusted model 1 and model 2. Each judgment about the risk of bias in a study is summarized by the index $R_j$ which takes binary values 0 (no bias) or 1 (bias). In bias-adjusted model 1, we extend the method introduced by Dias et al. (24) by adding a treatment-specific bias term $\gamma_{2,jbk}R_j$ to the relative treatment effect for both the AD and IPD parts of the model (40). A multiplicative model can also be employed, where treatment effects are multiplied by $\gamma_{1,jbk}^{R_j}$. These bias terms penalize the high RoB studies for potential overestimation or underestimation by adjusting their relative treatment effects. Next, we extend the model from Section 3.1 to adjust for bias.

**Part I: NMR model for IPD studies**

We model the IPD studies from both designs simultaneously; we differentiate between the designs by including the study-level bias terms. We can add either multiplicative $\gamma_{1,jbk}$ bias effects, additive $\gamma_{2,jbk}$ bias effects, or both (in this case, $\delta_{jbk}$ should be dropped from the additive part) as

$$\text{logit}(p_{ijk}) = \begin{cases} u_{jb} + \beta_{0j}x_{ijk} & \text{if } k = b \\ u_{jb} + \overbrace{\delta_{jbk}\gamma_{1,jbk}^{R_j}}^{multiplicative} + \overbrace{\delta_{jbk} + \gamma_{2,jbk}R_j}^{additive} + & \text{if } k \neq b \\ \beta_{0j}x_{ijk} + \beta_{1,jbk}^W x_{ijk} + (\beta_{1,jbk}^B - \beta_{1,jbk}^W)\bar{x}_j. \end{cases} \quad (1)$$

where $j = 1, \dots, ns_{IPD,RCT} + ns_{IPD,NRS}$.

The bias indicator $R_j$ follows a Bernoulli distribution with a bias probability $\pi_j = P(R_j = 1)$

$$R_j = \begin{cases} 1, & \text{if study } j \text{ has high risk of bias} \\ 0, & \text{otherwise} \end{cases}$$

$$R_j \sim \text{Bernoulli}(\pi_j).$$

Then based on the risk of bias for each study, a different beta distribution is placed for $\pi_j$.

$$\pi_j \sim Beta(a_1, a_2).$$

The hyperparameters $a_1$ and $a_2$ should be chosen in a way that reflects the risk of bias for each study. The degree of skewness in beta distribution can be controlled by the ratio $a_1/a_2$. When $a_1/a_2$ equals 1 (or $a_1 = a_2$), there is no skewness in the beta distribution (the distribution is reduced to a uniform distribution), which is appropriate for studies with unclear risk of bias. When the ratio $a_1/a_2$ is closer to 1, the more the mean of probability of bias (expected value of $\pi_j = a_1/(a_1 + a_2)$) gets closer to 1 and the study acquires 'major' bias adjustment.

Alternatively, we can use the study characteristics $\mathbf{z_j} = (z_{1,j}, z_{2,j}, \dots, z_{m,j})$ (eg, the concealment of the study) to predict $\pi_j$ through a logistic transformation as follows

$$logit(\pi_j) = e + \mathbf{f}^T \mathbf{z_j}.$$

where $\mathbf{f}^T = (f_1, \dots, f_m)$ is a vector of covariate effect on the odds ratio of bias and $e$ is the overall odds of bias. The superscript **T** transposes the vector. A minimally informative prior is located in the regression coefficients $e, \mathbf{f}^T \sim N(0, 10^2)$.

We alternatively describe the logistic model with additive bias effect in equation (1) by the following parametrisation

$$\text{logit}(p_{ijk}) = \begin{cases} u_{jb} + \beta_{0j}x_{ijk} & \text{if } k = b \\ u_{jb} + (1 - R_j)\delta_{jbk} + R_j\delta_{jbk}^{bias} + & \text{if } k \neq b \\ \beta_{0j}x_{ijk} + \beta_{1,jbk}^W x_{ijk} + (\beta_{1,jbk}^B - \beta_{1,jbk}^W)\bar{x}_j. \end{cases} \quad (2)$$

where $\delta_{jbk}^{bias} = \delta_{jbk} + \gamma_{jbk}$.

**Part II: NMR model for AD studies**

Similarly, we add the two bias terms to model the summary data.

$$\text{Logit}(p_{.jk}) = \begin{cases} u_{jb} & \text{if } k = b \\ u_{jb} + \overbrace{\delta_{jbk}\gamma_{1,jbk}^{R_j}}^{multiplicative} + \overbrace{\delta_{jbk} + \gamma_{2,jbk}R_j}^{additive} + \beta_{1,jbk}^B \bar{x}_j & \text{if } k \neq b \end{cases} \quad (3)$$

where $j = ns_{IPD,RCT} + ns_{IPD,NRS} + 1, \ldots, ns_{IPD,RCT} + ns_{IPD,NRS} + ns_{AD,RCT} + ns_{AD,NRS}$. Again, when multiplicative and additive parts are both considered in the model, the term $\delta_{jbk}$ needs to be removed from the additive term.

Other parametrisation of the logistic model with additive bias effect in equation (3) is

$$\text{Logit}(p_{.jk}) = \begin{cases} u_{jb} & \text{if } k = b \\ u_{jb} + (1 - R_j)\delta_{jbk} + R_j\delta_{jbk}^{bias} + \beta_{1,jbk}^B \bar{x}_j & \text{if } k \neq b. \end{cases} \quad (4)$$

**Part III: Combine the evidence from IPD and AD**

In addition to the covariates' effects and the treatment effects, here we also combine the multiplicative and the additive treatment-specific bias effects across studies by assuming they are either exchangeable ($\gamma_{1,jbk} \sim N(g_{1,bk}, \tau_{1,\gamma}^2)$, $\gamma_{2,jbk} \sim N(g_{2,bk}, \tau_{2,\gamma}^2)$) or fixed ($\gamma_{1,jbk} = g_{1,bk}$ and $\gamma_{2,jbk} = g_{2,bk}$). We set priors for the between-study standard deviation again as $\tau_{1,\gamma}, \tau_{2,\gamma} \sim \text{Unif}(0,2)$.

For the other parameterisation in equations (2) and (4), the bias-adjusted relative treatment effect $\delta_{jbk}^{bias}$ can be assumed exchangeable across studies

$$\delta_{jbk}^{bias} \sim N\left(g_{bk} + d_{Ak} - d_{Ab}, \frac{\tau^2}{q_j}\right)$$

or fixed as

$$\delta_{jbk}^{bias} = g_{bk} + d_{Ak} - d_{Ab}.$$

In this case, instead of assigning prior to the between-study heterogeneity in bias effect $\tau_\gamma$, we model the RoB weight $q_j = \tau^2/(\tau^2 + \tau_\gamma^2)$ for each study. The quantity represents the proportion of the between-study heterogeneity that is not explained by accounting for risk of bias. These weights take values between 0 and 1, $0 < q_j < 1$, and they are either given fixed values (as Spiegelhalter and Best proposed (41)) or assigned a prior to let the data estimate them, $q_j \sim Beta(v, 1)$ (as Verde assumed (25)). The values of $v$ determine the extent studies at high risk of bias will be down-weighted on average. Setting $v = 1$ gives $E(q_j) = v/(v + 1) = 0.5$, which means that high risk of bias studies will be penalized by 50% on average.

Dias et al. (24) proposed to model the mean bias effect $(g_{1,bk}, g_{2,bk})$ based on the compared treatments. One approach is to assume a common mean bias for studies that compare active treatments with an inactive treatment (placebo, standard, or no treatment)

$$g_{m,bk} = \begin{cases} g_m & \text{if } b \text{ is inactive treatment} \\ 0 & \text{if } b \text{ and } k \text{ are active treatments} \end{cases}$$

where $m = \{1,2\}$.

In this case, the mean bias effect cancels out contrasts for comparing two active treatments. When exchangeable bias parameters are used, active vs active comparisons have an expected bias effect of zero with uncertainty the common bias-heterogeneity parameters $\tau_{1,\gamma}^2$, $\tau_{2,\gamma}^2$ for multiplicative and additive, respectively.

Instead of assuming zero bias in active vs active comparison, we could assume a common and fixed bias effect $g_m^{act}$:

$$g_{m,bk} = \begin{cases} g_m & \text{if } b \text{ is inactive treatment} \\ (-1)^{dir_{bk}} g_m^{act} & \text{if } b \text{ and } k \text{ are active treatments} \end{cases}$$

The direction of bias ($dir_{bk}$) varies by the comparison type and should be defined in the data. The bias in active vs inactive comparisons will favour the active treatment. However, the direction of bias is less clear in studies that compare active treatments with each other. The direction of bias could be linked to other types of bias, such as 'optimism bias'–a bias favouring the newest treatment. In this case, the direction of bias in each active vs active comparison is set to be either 0, meaning that bias favours $b$ over $k$; or 1, meaning that $k$ is favoured to $b$. We could also follow a data-driven approach and assign the bias direction a Bernoulli distribution $dir \sim \text{Bernoulli}(p_{dir})$ where the probability of $b$ to be favoured over $k$, $p_{dir}$, is given a beta distribution $p_{dir} \sim Beta(a_3, a_4)$. The shape of this beta distribution is characterized by $a_3$ and $a_4$. When $a_3$ is set a value less than $a_4$, the study is more likely to be favouring $b$ over $k$.

### *3.2.4 Bias-adjusted model 2*

Extending the model initially introduced by Verde (25), bias-adjusted model 2 parametrises the relative treatment effect using a bimodal normal distribution that involves the bias parameters (25). We define the bias-adjusted relative treatment effect $\theta_{jbk}$ as follows in both parts of the NMR.

**Part I: NMR model for IPD studies**

$$\text{Logit}(p_{ijk}) = \begin{cases} u_{jb} + \beta_{0j}x_{ijk} & \text{if } k = b \\ u_{jb} + \theta_{jbk} + \beta_{0j}\,x_{ijk} + & \text{if } k \neq b \\ \beta^W_{1,jbk}x_{ijk} + (\beta^B_{1,jbk} - \beta^W_{1,jbk})\,\bar{x}_j. \end{cases}$$

**Part II: NMR model for AD studies**

We also add the bias adjustment term to AD part

$$Logit(p_{.jk}) = \begin{cases} u_{jb} & \text{if } k = b \\ u_{jb} + \theta_{jbk} + \beta^B_{1,jbk}\,\bar{x}_j & \text{if } k \neq b. \end{cases}$$

**Part III: Combine the evidence from IPD and AD**

The coefficients from the covariates effect and treatment effects are combined as in the previous models. We additionally combine the bias-adjusted relative treatment effect $\theta_{jbk}$ via exchangeable model with a mixture of two normal distributions

$$\theta_{jbk} \sim (1 - \pi_j)N(d_{Ak} - d_{Ab}, \tau^2) + \pi_j N(d_{Ak} - d_{Ab} + \gamma_{jbk}, \tau^2 + \tau_\gamma^2).$$

Assuming a fixed-effect model we can alternatively summarize these relative effects

$$\theta_{jbk} = (1 - \pi_j)(d_{Ak} - d_{Ab}) + \pi_j(d_{Ak} - d_{Ab} + \gamma_{jbk}) = d_{Ak} - d_{Ab} + \pi_j\gamma_{jbk}.$$

This model adjusts the relative treatment effect by a bias effect that is proportional to the bias probability in each study. The bias parameters $\gamma_{jbk}$ across studies are assigned either the exchangeable- or fixed-effect model and then the mean bias effects $g_{bk}$ are also combined across comparisons.

Following what we describe in Section 3.2.3, the between-study standard deviation $\tau_\gamma$ can also be modelled in two different ways. We set a prior either for $\tau_\gamma \sim \text{Unif}(0,2)$ or for $q_j \sim Beta(v, 1)$ where $q_j = \tau^2/(\tau^2 + \tau_\gamma^2)$ represents the RoB weight for each study. However, choosing the prior for $q_j$ could be more meaningful in practice as $v$ represents the discounting in study weight. All other syntheses are performed as outlined for bias-adjusted model 1 in Section 3.2.3.

# 4 Implementation of the models and results

We implemented the models in a Bayesian setting using Just Another Gibbs Sampler (42) software through R (43). For all models, we ran two chains each for 100 000 iterations, discarded the first 40 000 samples, and thinned by 1. We examined the convergence of chains on each parameter by either visually inspecting the trace plots or checking the Gelman-Rubin statistic, $\hat{R}$, which measures the agreement between the within- and between-chains of MCMC; it should be approximately 1 when the chain converges properly. From here onwards, point estimates refer to posterior medians.

## *4.1 Immunomodulatory agents in RRMS*

We conducted NMA and NMR assuming a common treatment effect across studies (the small number of studies did not allow efficient estimation of heterogeneity). We included age as a covariate in the NMR model which was centred around mean age 38 to improve convergence. We also assumed a common age effect across studies. For the IPD part of the models, we set the within- and between-study age effects equal: $\beta^W_{1,jbk} = \beta^B_{1,jbk}$. In bias-adjusted models 1 and 2, we assigned two different informative prior distributions for the bias probability $\pi_j$: a $Beta(100,1)$ for high RoB studies and $Beta(1,100)$ for low RoB studies (see Appendix Figure 2 and Table 3). We assumed additive bias effects and combined them across studies into a common parameter. The direction of bias was assumed to favour the active treatment rather placebo in RCTs and any other treatment over glatiramer acetate in the SMSC since it is the oldest treatment. We set placebo as a network reference for all analyses except when using NRS information as a prior; in that case, natalizumab was used as the reference treatment.

We first analysed the data using the SMSC data to construct priors for the treatment effects. The posterior distributions of the logORs were $d_{DF\ vs\ N} \sim N(-0.01, 0.2)$ (dimethyl fumarate vs natalizumab) and $d_{GA\ vs\ N} \sim N(1.56, 0.33)$ (glatiramer acetate vs natalizumab). The basic parameter of placebo vs natalizumab (not observed in the cohort) was assigned an approximately uninformative prior ($d_{P\ vs\ N} \sim N(0, 10^2)$). In Appendix Figure 1, we present the results when these posteriors were used as (discounted) priors in the NMA of the RCT data assuming different values of $w$. Only the estimated effect of glatiramer acetate vs natalizumab changed slightly when incorporating the non-randomized evidence because the SMSC has a much smaller sample size ($n =206$) than all RCTs together ($n =3\ 891$).

Figure 2 and Appendix Table 4–7 show the NMA ORs and the corresponding 95% credible intervals (CrI) using no adjustment and bias-adjusted models 1 and 2. The adjustment for the different bias effects did not materially change the estimated ORs. The small change we observed for glatiramer acetate in bias-adjusted models can be attributed to the high risk of bias in Bornstein and Johnson studies (32,33).

For bias-adjusted model 1, the bias effect $exp(g)$ was estimated 0.705 (95% CrI: 0.198–1.459). The OR of the active treatments when compared with placebo in low RoB studies are on average 0.705 times the OR in low RoB studies, yet the uncertainty is very large. In the bias-adjusted model 2, $exp(g)$ was more precisely estimated at 0.323 (95% CrI: 0.126–0.821). This means that on average high RoB studies tend to overestimate the efficacy of the active treatments. We investigated the convergence of the model parameters in Appendix Figure 4 and Appendix Table 8. The bias parameter $g$ estimated from the bias-adjusted model 1 has a slightly poor convergence when compared with other parameters.

We incorporated the effect of age in bias-adjusted model 1; Figure 3 presents the NMR ORs of active vs placebo for various age values. The estimated age coefficient $exp(B)$ was 0.984 (95% CrI: 0.264–1.935) suggesting that the ORs of each treatment vs placebo increases with age.

## *4.2 Antidepressants for major depression*

We conducted an NMA assuming a random treatment effect across studies. For bias-adjusted models 1 and 2, we used additive bias effects and combined them across studies assuming random-effects. The bias probability $\pi_j$ of moderate and low RoB studies was given prior distributions $Beta(20,1)$ and $Beta(1,20)$, respectively (see Appendix Figure 3). When we set the direction of bias in studies comparing an active drug to placebo, we assumed mean bias $g^P$, and the antidepressant was assumed the favoured treatment; then in active vs active comparisons, we assumed bias $g^{act}$, and the sponsored treatment was assumed the favoured treatment. In other cases, the mean bias was set to zero. We performed a sensitivity analysis to investigate the robustness of the results with less informative prior distributions for the bias probability $\pi_j$ in both bias models $Beta(10,1)$ and $Beta(1,10)$ for studies at moderate and low RoB, respectively.

Table 4 shows the estimates of bias effect parameters using the bias-adjusted models 1 and 2. The results suggest that moderate RoB studies do not provide different estimates of the effectiveness of the active interventions versus placebo, whereas the effects of sponsored treatments are overestimated on average. We also fitted the bias-adjusted models 1 and 2 by re-parametrising the heterogeneity $\tau_\gamma$ using the weights $q_j$. We set $q_j = 1$ for studies at low RoB and $q_j \sim Beta(1/3,1)$ for moderate RoB studies which reduces their weight on average by 25% or $q_j \sim Beta(4,1))$ for 80% weight reduction. The results do not materially change.

Figure 4 presents the resulting OR and 95% CrI for the adjusted and unadjusted models. Controlling for the information from the moderate RoB studies scarcely changed the effects of active drugs vs placebo. Using less informative priors for the bias probability did not materially change these conclusions (Appendix Figure 5). The estimate of between-study heterogeneity in treatment effect was 0.210 (95% CrI: 0.169–0.251) in unadjusted model, which decreased when bias-adjusted model 1 was applied to 0.176 (95% CrI: 0.089–0.236) and the estimate in bias-adjusted model 2 was 0.213 (95% CrI: 0.147–0.291).

# 5 Discussion

We introduced a suite of Bayesian NMA and NMR models to synthesize evidence that comes from different study designs and in different data formats. We extended the three-level hierarchical model for combining IPD and AD with four models incorporating RCT and NRS evidence. The first model ignores differences in design and RoB between studies; the second uses NRS to construct discounted treatment effect priors; and two models adjust for the risk of bias in each study. We implemented the four NMA/NMR models in a dataset comparing treatments for RRMS patients. The estimated treatment effects were consistent, irrespective of the model used. When age was included as a covariate, the efficacy of active treatments relative to placebo decreased with increasing age. In other words, all active treatments become less effective for older patients, which aligns with previous findings (44,45).

We also illustrated the bias-adjusted models in a network of AD from RCTs on antidepressants. The results from sponsored drug arms in head-to-head studies tended to be larger than those in non-sponsored arms. In the original analysis, Cipriani et al. (38) did not detect any impact of sponsoring in the estimated efficacy of the antidepressants. Note, however, that our bias-adjusted models estimate the interaction between risk of bias and sponsoring and hence it is possible that sponsoring plays a role in modifying the treatment effect only in studies with moderate risk of bias.

Our methods tackle the bias issue at the quantitative synthesis stage. The bias (for NRS, in particular) should also be mitigated at study design and when interpreting results. In their comprehensive framework, Sarri et al. (46) proposed seven steps outlining how to combine RCT and NRS data in NMA. They proposed different considerations for interpreting findings, suggesting a way that reflects the differences in evidence type. Their framework suggests a certain critical assessment of NRS, which can be used in our bias-adjusted models.

Some limitations of our proposed models need to be acknowledged. First, the bias-adjusted models require several studies at different levels of RoB. In the absence of many studies, strong assumptions can be imposed on bias parameters via informative priors. Our first example of RRMS only included six studies; we assigned highly informative beta distributions to the bias probability. We used less informative priors in the case of antidepressants' network because many studies were available. Second, the results of the analysis can be sensitive to the prior assumptions in model parameters. For this reason, sensitivity analyses should be conducted to investigate the robustness of the estimates, using different priors if possible. Sensitivity to prior distributions is particularly important for the probability of bias and the covariate effect parameters. In our examples, we found that bias-adjusted model 2 was more sensitive to the prior assigned to the bias probability when compared with bias-adjusted model 1. Finally, choosing down-weighting parameters for the model that uses the NRS data to construct prior information is not straightforward. However, Efthimiou et al. (23) outlined different considerations to guide this choice.

To implement the proposed models, there are further worthy considerations. These include performing a comprehensive systematic review to identify relevant RCTs and NRSs (following the framework introduced by Sarri et al. (46)). In our RRMS example, we included RCTs identified in a previous systematic review with available IPD (27,28) and observational data from the SMSC. For clinically-relevant results after analysis, more data needs to be included to apply our methods to an extended network of all drugs used to treat patients with RRMS, such as presented by Jenkins et al. (47). In their review, Jenkins et al. showed how including NRS data in the synthesis model increased the between-study heterogeneity and therefore the uncertainty around the effect estimates. By accounting for potential effect modifiers and differences in RoB, other studies can investigate whether our models explain large between-study heterogeneity.

Combining individual and aggregate data has two key advantages when compared with analysing aggregate data solely. First, aggregate data studies contribute only to estimating interactions between mean values of effect modifiers and treatment, yet individual data studies account for interactions at the individual patient-level, thus avoiding ecological bias. Second, individual data adjust for prognostic factors and covariates that predict the outcome and the course of the disease regardless of the assigned treatment (44). Adjusting for prognostic factors is desirable (48) in order to improve the interpretation and the external validity of the findings (49); enhance the precision of the estimated treatment effects (50); and correct potential imbalance in baselines after randomisation (51).

Incorporating NRS evidence into NMA models that traditionally only include RCTs is increasingly important in several clinical research settings, such as when conducting RCTs are less feasible for rare conditions. A recent scoping review of methods that combine RCT and NRS in NMA (52) reveals that unadjusted synthesis is the most popular approach, probably for its ease of use. The unadjusted analysis, however, can be considered as an initial step but not the primary analysis, as it ignores the differences in design and RoB. Accounting for within-study bias in both observational and experimental data, our suite of models offers a viable alternative. Our approach also allows estimating individualized treatment effects through the inclusion of participant characteristics.

# Tables

*Table 1 Notation for the synthesis models*

| Notation | Description |
|---|---|
| $i = 1, \ldots, np_j$ | participant id |
| $j = 1, \ldots, ns$ | study id |
| $k = 1, \ldots, K$ | treatment index |
| $ns_{IPD}, ns_{AD}, ns_{RCT}, ns_{NRS}$ $ns_{IPD,RCT}, ns_{AD,RCT}$ $ns_{IPD,NRS}, ns_{AD,NRS}$ | the number of studies. The index refers to the design or format of the study or both |
| $y_{ijk}$ | binary outcome (0/1) |
| $p_{ijk}$ | probability of the event to occur |
| $r_{jk}$ | the number of events per arm |
| $n_{jk}$ | the sample size per arm |
| $b$ | the study-specific reference |
| $u_{jb}$ | The treatment effect of the study-specific reference $b$ when $x_{ijk} = \bar{x}_j = 0$ |
| $\delta_{jbk}$ | log(OR) of treatment k relative to $b$ |
| $x_{ijk}$ | the covariate |
| $\bar{x}_j$ | the mean covariate for study j |
| $d_{Ak}$ | the basic parameters where $d_{AA}$=0 when A set as the reference in the network |
| $z_j$ | study characteristics to estimate the bias probability $\pi_j$ |

*Table 2 Assumptions about the model parameters*

| Parameter | Assumptions |
|---|---|
| Relative treatment effect ($\delta_{jbk}$) | Random-effects: $\delta_{jbk} \sim N(d_{Ak} - d_{Ab}, \tau^2)$ |
| | Common-effect: $\delta_{jbk} = d_{Ak} - d_{Ab}$ |
| Covariate effect ($\beta_{0j}$) | Independent effects: $\beta_{0j} \sim N(0, 10^2)$ |
| | Random-effects: $\beta_{0j} \sim N(B_0, \tau_0^2)$ |
| Within-study covariate-treatment interaction ($\beta_{1,jbk}^W$) | Random-effects: $\beta_{1,jbk}^W \sim N(B_{1,Ak}^W - B_{1,Ab}^W, \tau_W^2)$ |
| | Common-effect: $\beta_{1,jbk}^W = B_{1,Ak}^W - B_{1,Ab}^W$ |
| Between-study covariate-treatment interaction ($\beta_{1,jbk}^B$) | Random-effects: $\beta_{1,jbk}^B \sim N(B_{1,Ak}^B - B_{1,Ab}^B, \tau_B^2)$ |
| | Common-effect: $\beta_{1,jbk}^B = B_{1,Ak}^B - B_{1,Ab}^B$ |
| Bias effect ($\gamma_{m,jbk}$) $m = \{1,2\}$ | Random-effects: $\gamma_{m,jbk} \sim N(g_{m,bk}, \tau_{m,\gamma}^2)$ |
| | Common-effect: $\gamma_{m,jbk} = g_{m,bk}$ |
| Mean bias effect ($g_{m,bk}$) | $g_{m,bk} = \begin{cases} g_m & \text{if } b \text{ is inactive treatment} \\ 0 & \text{if } b \text{ and } k \text{ are active treatments} \end{cases}$ |
| | $g_{m,bk} = \begin{cases} g_m & \text{if } b \text{ is inactive treatment} \\ (-1)^{dir_{bk}} g_m^{act} & \text{if } b \text{ and } k \text{ are active treatment} \end{cases}$ |
| Bias indicator | $R_j \sim Bernoulli(\pi_j)$ |
| Bias probability ($\pi_j$) | $\pi_j \sim Beta(a_1, a_2)$ |
| | $logit(\pi_j) = e + \mathbf{f}^T \mathbf{z}_j$ |

*Table 3 Study characteristics and assigned priors for bias probability*

*IPD; Individual Participant Data, AD; Aggregate Data, RCT; Randomized Controlled Trial,*

*NRS; Non-Randomised Study*

| Study | Treatments | Number of patients with at least one relapse in two years | Sample size | Design & data formal | Risk of bias (RoB) | Mean age | Distribution of bias probability $\pi_j$ |
|---|---|---|---|---|---|---|---|
| AFFIRM (29) | Natalizumab, Placebo | 359 | 939 | RCT IPD | low | 36 | Beta(1,100) |
| CONFIRM (30) | Dimethyl fumarate, Glatiramer acetate, Placebo | 451 | 1417 | RCT IPD | low | 37 | Beta(1,100) |
| DEFINE (31) | Dimethyl fumarate, Placebo | 394 | 1234 | RCT IPD | low | 39 | Beta(1, 100) |
| Swiss Multiple Sclerosis Cohort (34) | Dimethyl fumarate, Glatiramer acetate, Natalizumab | 44 | 206 | NRS IPD | high | 46 | Beta(100,1) |
| Bornstein (32) | Glatiramer acetate, Placebo | 30 | 50 | RCT AD | high | 34 | Beta(100,1) |
| Johnson (33) | Glatiramer acetate, Placebo | 186 | 251 | RCT AD | high | 30 | Beta(100,1) |

*Table 4 Results (estimates and 95% credible intervals) from bias-adjusted models 1 and 2 for the antidepressants network shown in Figure 1b*

$g^p$, the bias effect for active-placebo comparisons; $g^{act}$, the bias effect for active-active comparisons (sponsored treatment assumed to be favoured); RoB, risk of bias in the study; CrI, credible interval.

| | Bias-adjusted model 1 | Bias-adjusted model 2 |
|---|---|---|
| Model assuming a prior $\tau_\gamma \sim \text{Unif}(0,2)$ for the heterogeneity in bias effects | | |
| **Primary analysis: Bias probability distribution (low RoB: $\pi_j \sim \text{Beta}(1,20)$, moderate RoB: $\pi_j \sim \text{Beta}(20,1)$)** | | |
| Mean bias effect: $\exp(g^p)$ | 1.090 (0.975, 1.249) | 1.035 (0.939, 1.143) |
| Mean bias effect: $\exp(g^{act})$ | 1.186 (1.054, 1.335) | 1.182 (1.054, 1.335) |
| Heterogeneity in bias effect: $\tau_\gamma$ (95% CrI) | 0.130 (0.005, 0.261) | 0.185 (0.128, 0.251) |
| **Sensitivity analysis: Bias probability distribution (low RoB: $\pi_j \sim \text{Beta}(1,10)$, moderate RoB; $\pi_j \sim \text{Beta}(10,1)$)** | | |
| Mean bias effect: $\exp(g^p)$ | 1.163 (0.966, 1.421) | 1.035 (0.878, 1.224) |
| Mean bias effect: $\exp(g^{act})$ | 1.257 (1.095, 1.478) | 1.271 (1.094, 1.600) |
| Heterogeneity in bias effect: $\tau_\gamma$ | 0.206 (0.078, 0.318) | 0.210 (0.127, 0.354) |
| Model that re-parametrises the heterogeneity using weights $q_j \sim Beta(v,1)$ where $q_j = \tau^2/(\tau^2 + \tau_\gamma^2)$ | | |
| **Low RoB studies: no down-weighting; Moderate RoB studies: down-weight by 25%** | | |
| Mean bias effect: $\exp(g^p)$ | 0.985 (0.786, 1.475) | 0.817 (0.549, 1.112) |
| Mean bias effect: $\exp(g^{act})$ | 1.222 (1.073, 1.476) | 1.427 (1.173, 1.942) |
| **Low RoB studies: no down-weighting; Moderate RoB studies: down-weighting by 80%** | | |
| Mean bias effect: $\exp(g^p)$ | 1.012 (0.860, 1.167) | 1.008 (0.851, 1.153) |
| Mean bias effect: $\exp(g^{act})$ | 1.203 (1.067, 1.383) | 1.231 (1.081, 1.470) |

**Figures**

(a)

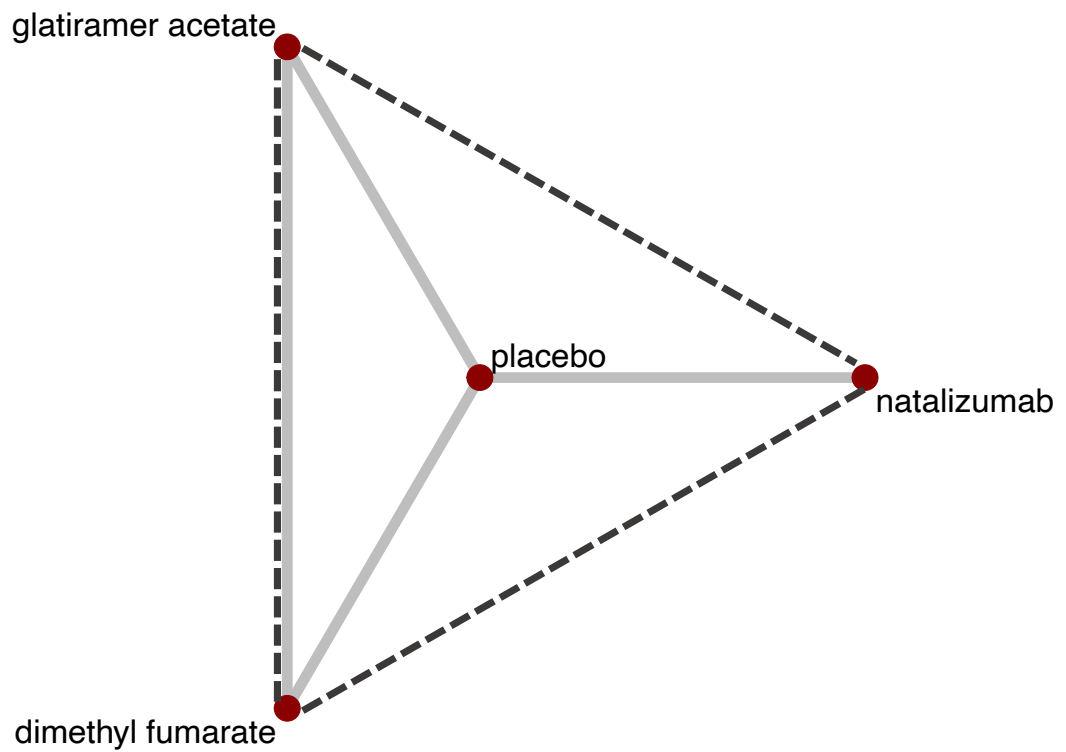

(b)

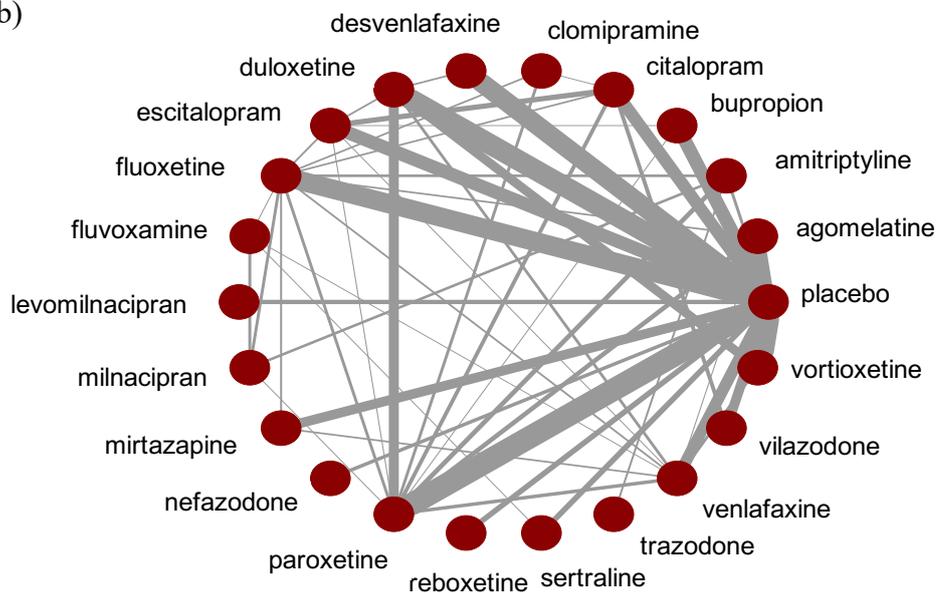

*Figure 1 Network plots of (a) treatments for patients with relapsing-remitting multiple sclerosis compared in randomised controlled trials (solid, grey edges) and in the Swiss Multiple Sclerosis Cohort (dashed, black edges). The outcome is relapse in two years (b) antidepressants and placebo compared in randomised controlled trials. The outcome is response to treatment. The thickness of the edges is proportional to the number of trials comparing each pair of treatments.*

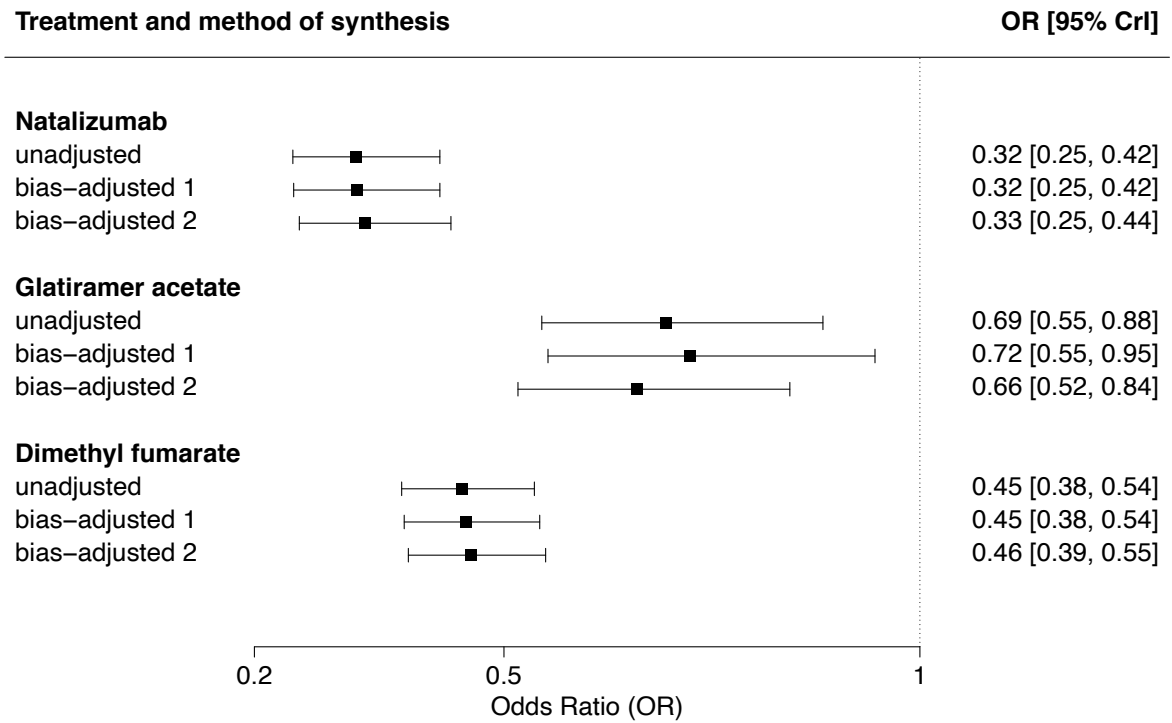

*Figure 2 Relapse odds ratios with 95% credible intervals of active treatments vs placebo among patients with relapsing-remitting multiple sclerosis. The estimates are computed by conducting unadjusted analysis and bias-adjusted analysis 1 and 2 in the data in the network of Figure 1a.*

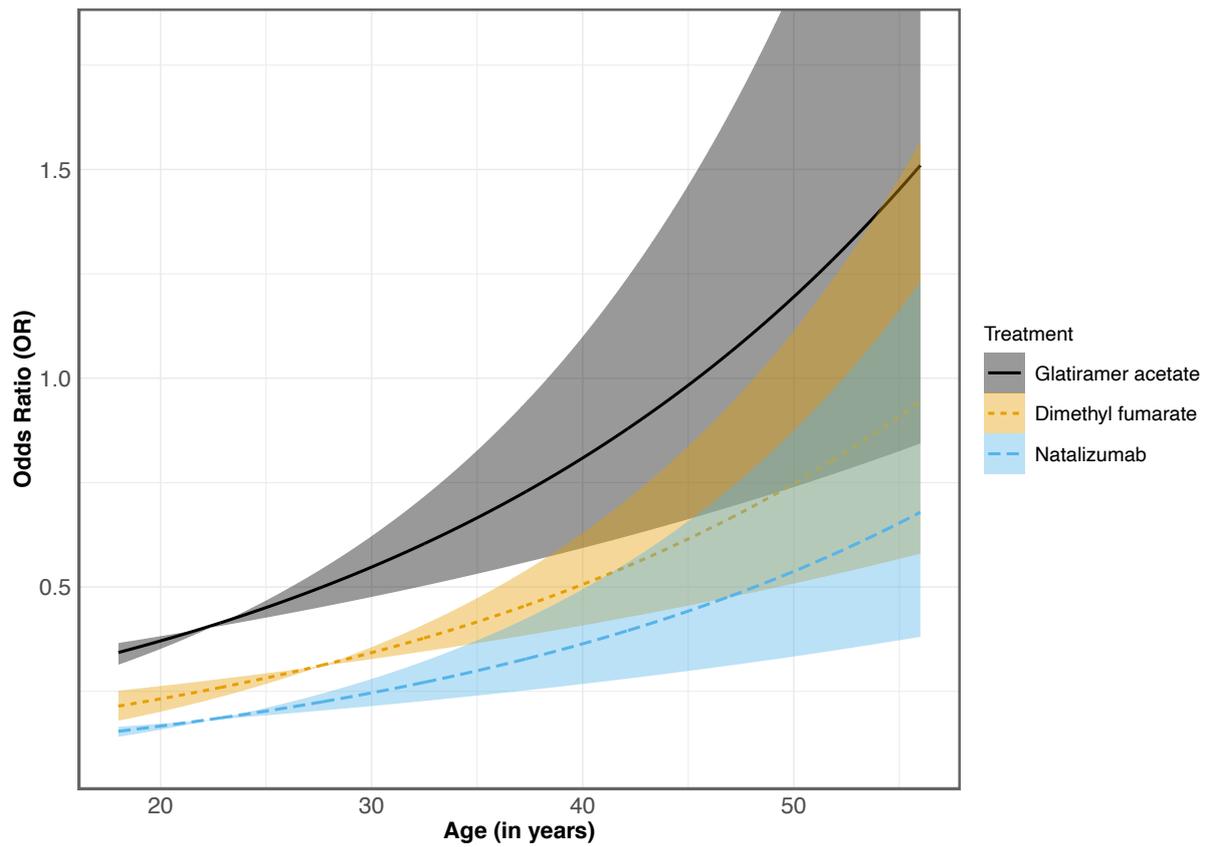

*Figure 3 The relationship between patient age (in years) and the estimated odds ratio with 95% credible intervals (the shaded areas) for active treatments vs placebo among patients with relapsing-remitting multiple sclerosis estimated with network meta-regression with bias-adjusted model 1.*

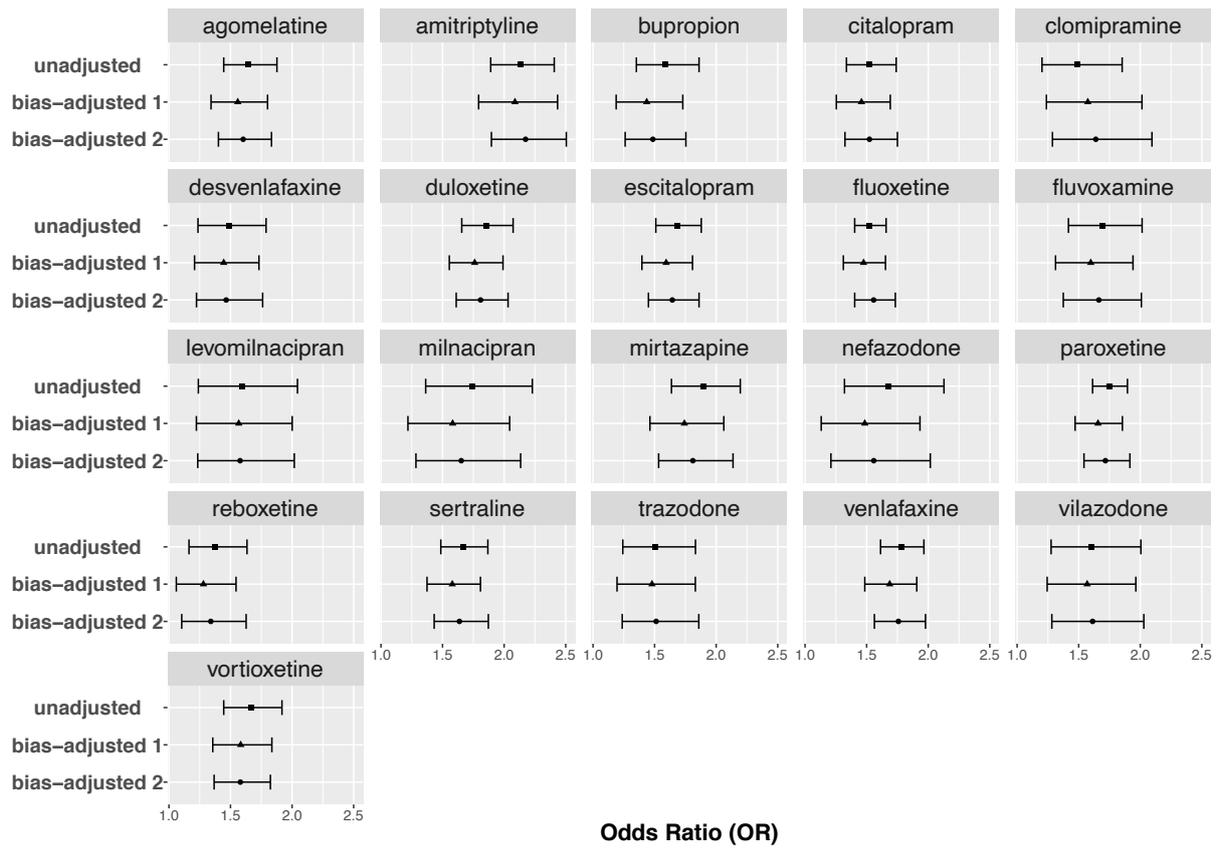

*Figure 4 Response odds ratio with 95% credible interval for each antidepressant vs placebo estimated from unadjusted analysis and bias-adjusted models 1 and 2 using the data presented in the network of Figure 1b. A random-effects network meta-analysis model is assumed to estimate treatment and bias effects.*